\documentclass[reqno, centertags]{amsart}
\usepackage{amssymb}

\newcommand{\ket}[1]{{|#1\,\rangle}}

\newcommand{\C}[1]{{\mathcal{#1}}} 
\newcommand{\D}[1]{{\mathbb#1}}  
\theoremstyle{plain}
\newtheorem{theorem}{Theorem}[section]
\newtheorem{lemma}{Lemma}[section]

\theoremstyle{definition}
\newtheorem{definition}{Definition}[section]

\theoremstyle{remark}
\numberwithin{equation}{section}

\def\nl{\par\noindent}
\def\part #1{\left(#1\right)}
  \def\parq #1{\left[#1\right]}
  \def\parg #1{\left\{#1\right\}}
  \def\para #1{\left\langle#1\right\rangle}
  \def\c2{\D C^2}
  
  \def\tensorn{\otimes^n\D C^2}
    \def\tensorm{\otimes^m\D C^2}
\def\kz{\ket{0}}
\def\ko{\ket{1}}

\def\kj{\ket{j}}
\def\kk{\ket{k}}
\def\kpsi{\ket{\psi}}
\def\kphi{\ket{\varphi}}
\def\sumn{\sum_{j=0}^{2^n-1}}
\def\summ{\sum_{k=0}^{2^m-1}}
\def\cpphi{C^+{\ket{\varphi}}}
\def\cppsi{C^+{\ket{\psi}}}
\def\cmphi{C^-{\ket{\varphi}}}

\def\sumppsi{\sum_{j\in\cppsi}}

\def\kphin{\sumn a_j\kj}

\def\AND{{\tt AND}}
\def\NOT{{\tt NOT}}
\def\OR{{\tt OR}}
\def\SQR{\sqrt{{\tt NOT}}}

\def\P{{\rm Prob}}
\def\sqr{\sqrt{\neg}}

  \def\norm#1{\Vert #1\Vert}
\newcommand{\IC}{{\mathbb C}} 
\newcommand{\parto}[1]{\left(#1\right)}
   
\def\mand{\curlywedge}
\def\mor{\curlyvee}

\begin{document}

\title{An unsharp  logic from quantum
computation}

\author[G.~Cattaneo]{Gianpiero Cattaneo}
\address[G.~Cattaneo]{
Dipartimento di Informatica, Sistemistica e Comunicazione,
 Universit\`a di Milano - Bicocca,
          Via Bicocca degli Arcimboldi 8,
          I-20126 Milano (Italy)
}
\email{cattang@disco.unimib.it}
\author[M.~L.~Dalla Chiara]{Maria Luisa Dalla Chiara}
\address[M.~L.~Dalla Chiara]{Dipartimento di Filosofia,
      Universit\`a di Firenze,
      Via Bolognese 52,  I-50139 Firenze (Italy)}

\email{dachiara@mailserver.idg.fi.cnr.it}

\author[R.~Giuntini]{Roberto Giuntini}
\address[R.~Giuntini]{Dipartimento di Scienze Pedagogiche e Filosofiche,
   Universit\`a di Cagliari,
   Via Is Mirrionis 1,  I-09123 Cagliari (Italy)}
\email{giuntini@unica.it}
\author[R.~Leporini]{Roberto Leporini}
\address[R.~Leporini]{
Dipartimento di Informatica, Sistemistica e Comunicazione,
 Universit\`a di Milano - Bicocca,
          Via Bicocca degli Arcimboldi 8,
          I-20126 Milano (Italy)
}
\email{leporini@disco.unimib.it}
  \maketitle
\begin{abstract}
Logical gates studied in quantum computation suggest a natural
logical
 abstraction that gives rise to a new form of  {\it unsharp quantum logic\/}.
  We study the logical connectives corresponding to the following gates:
   the {\it Toffoli gate\/}, the $\NOT$ and the $\SQR$
   (which admit of natural physical models).
   This leads to a semantic characterization of a logic that we call
   {\it computational quantum logic\/} ($CQL$).
\end{abstract}

\section{Introduction}
The theory of quantum computation naturally suggests the semantic
characterization for a new form of quantum logic, that turns out
to have some typical  {\it unsharp \/} features. According
to this semantics, the {\em meaning\/} of a sentence is identified
with a {\em system of qubits\/}, a vector belonging to a
convenient Hilbert space, whose dimension depends on the logical
complexity of our sentence. At the same time, the {\em logical
connectives\/} are interpreted as particular {\em logical
gates\/}.

\section{Quantum logical gates}
We will first sum up some basic notions of quantum computation.

Consider the two-dimensional Hilbert space $\c2$, where any vector
$\ket{\psi}$ is represented by a pair of complex numbers.
 Let $B=\parg{\ket{0},\ket{1}}$ be the {\it orthonormal basis \/} for $\c2$
 such that $$\ket{0}= (0,1); \,\,\, \ket{1}=(1,0).$$

\begin{definition}
{\it Qubit} \label{de:qubit} \nl
A {\it qubit\/} is a unit vector
$\ket{\psi}$ of the space $\c2$.
\end{definition}
Hence, any {\it qubit\/} will have the following form: $$
\ket{\psi}= a_0\ket{0} + a_1\ket{1}, $$ where $a_0,a_1\in\D C$ and
$|a_0|^2+|a_1|^2=1$.

We will use $x,y,\ldots$ as variables ranging over the set
$\parg{0,1}$. At the same time, $\ket{x},\ket{y},\ldots$ will
range over the basis $\parg{\ket{0},\ket{1}}$. Further we will use
the following abbreviation: \nl
$\otimes^n\c2:=\underbrace{\c2\otimes\ldots\otimes\c2}_{n-times}$
(where $\otimes$ represents the tensor product).

The set of all vectors having the form
$\ket{x_1}\otimes\ldots\otimes\ket{x_n}$ represents an orthonormal
basis for $\otimes^n\c2$ (also called {\em computational
basis\/}). We will also write $\ket{x_1,\ldots,x_n}$ instead of
$\ket{x_1}\otimes\ldots\otimes\ket{x_n}$. \nl

\begin{definition}
{\it n-qubit system (or n-register)} \label{de:registro}\nl An
{\it n-qubit system \/} (or {\it n-register\/}) is any unit vector
$\ket{\psi}$ in the product space $\otimes^n\c2$.
\end{definition}

Apparently, the computational basis of $\tensorn$ can be labelled
by binary strings such as $$ \ket{\underbrace{011\ldots
10}_{n-times}}. $$

Since any string $\ket{\underbrace{011\ldots 10}_{n-times}}$
represents a natural number $j \in[0,2^n -1]$ in binary notation,
any unit vector of $\tensorn$ can be shortly expressed in the
following form:
$$
\sumn a_j\ket{j}.
$$

In the following we will call any vector that is either a qubit or
an $n$-qubit system a {\em quregister\/}. At the same time,
$\ket{0}$ and $\ket{1}$ will be also called {\em bits\/}.

We will now introduce some examples of {\it quantum logical
gates\/}. Generally, a quantum logical gate can be described as a
 unitary operator, assuming arguments and values in a
product-Hilbert space $\otimes^n\c2$. First of all we will study
the so called {\it Toffoli gate\/}. It will be expedient to start
by analysing the simplest case, where our Hilbert space has the
form:

$$ \otimes^3\c2=\c2\otimes\c2\otimes\c2. $$ In such a case, the
Toffoli gate will transform the vectors of $\otimes^3\c2$ into
vectors of $\otimes^3\c2$. In order to stress that our operator is
defined on the product space $\otimes^3\c2$, we will indicate it
by $T^{(1,1,1)}$. Since we want to define a unitary operator, it
will be sufficient to determine its behaviour for the elements of
the basis, having the form $\ket{x}\otimes\ket{y}\otimes\ket{z}$
(where $x,y,z \in
\parg{0,1}$).

\begin{definition}
{\it The Toffoli gate\/} $T^{(1,1,1)}$ \label{de:toffoli} \nl
The{\it Toffoli gate\/} $T^{(1,1,1)}$ is the linear operator
$T^{(1,1,1)}: \otimes^3\IC^2 \mapsto \otimes^3 \IC^2$ that is
defined for any element $\ket{x}\otimes\ket{y}\otimes\ket{z}$ of
the basis as follows: $$
T^{(1,1,1)}(\ket{x}\otimes\ket{y}\otimes\ket{z})
=\ket{x}\otimes\ket{y}\otimes\ket{\text{min}(x,y)
  \oplus z},
$$ where $\oplus$ represents the sum modulo $2$.
\end{definition}

>From an intuitive point of view, it seems quite natural to ``see"
the gate $T^{(1,1,1)}$ as a kind of ``truth-table" that transforms
triples of {\it zeros\/} and of {\it ones\/} into triples of {\it
zeros \/} and of {\it ones\/}. The ``table'' we will obtain is the
following:

\begin{align*}
\ket{0,0,0}\quad&\rightarrowtail\quad\ket{0,0,0}\\
\ket{0,0,1}\quad&\rightarrowtail\quad\ket{0,0,1}\\
\ket{0,1,0}\quad&\rightarrowtail\quad\ket{0,1,0}\\
\ket{0,1,1}\quad&\rightarrowtail\quad\ket{0,1,1}\\
\ket{1,0,0}\quad&\rightarrowtail\quad\ket{1,0,0}\\
\ket{1,0,1}\quad&\rightarrowtail\quad\ket{1,0,1}\\
\ket{1,1,0}\quad&\rightarrowtail\quad\ket{1,1,1}\\
\ket{1,1,1}\quad&\rightarrowtail\quad\ket{1,1,0}
\end{align*}

In the first six cases, $T^{(1,1,1)}$ behaves like the identity
operator;
 in the last two cases, instead, our gate transforms the last element of
 the triple into the opposite element ($0$  is transformed into  $1$ and $1$
 transformed into $0$).

One can easily show that $T^{(1,1,1)}$ has been well defined for
our aims: one is dealing with an operator that is not only linear
but also unitary. The matrix representation of $T^{(1,1,1)}$ is
the following: $$
  \begin{pmatrix}
    1 & 0 & 0 & 0 & 0 & 0 & 0 & 0 \\
       0 & 1 & 0 & 0 & 0 & 0 & 0 & 0\\
          0 & 0 & 1 & 0 & 0 & 0 & 0 & 0\\
             0 & 0 & 0 & 1 & 0 & 0 & 0 & 0 \\
                0 & 0 & 0 & 0 & 1 & 0 & 0 & 0\\
                   0 & 0 & 0 & 0 & 0 & 1 & 0 & 0\\
                      0 & 0 & 0 & 0 & 0 & 0 & 0 & 1\\
                         0 & 0 & 0 & 0 & 0 & 0 & 1 & 0
      \end{pmatrix}
$$

By using $T^{(1,1,1)}$, we can introduce a convenient notion of
{\it conjunction\/}. Our conjunction, which will be indicated by
{\tt AND}, is characterized as a function whose arguments are
pairs of vectors in $\IC^2$ and whose values are vectors of the
product
 space $\otimes^3\IC^2$.
\begin{definition}
{\tt AND} \label{de:and} \nl For any $\ket{\varphi}\in\IC^2$ and
any $\ket{\psi}\in \IC^2$: $$ {\tt
AND}\parto{\ket{\varphi},\ket{\psi}}:
=T^{(1,1,1)}\parto{\ket{\varphi}\otimes\ket{\psi}\otimes\ket{0}}.
$$

\end{definition}

Let us check that {\tt AND} represents a good generalization of
the corresponding classical truth-function. For the arguments
$\ket{0}$ and $\ket{1}$ we will obtain the following
``truth-table":
\begin{align*}
(\ket{0},\ket{0})\quad&\rightarrowtail\quad
T^{(1,1,1)}(\ket{0}\otimes\ket{0}\otimes\ket{0})=
                            \ket{0}\otimes\ket{0}\otimes\ket{0} \\
(\ket{0},\ket{1})\quad&\rightarrowtail\quad
T^{(1,1,1)}(\ket{0}\otimes\ket{1}\otimes\ket{0})=
                         \ket{0}\otimes\ket{1}\otimes\ket{0} \\
(\ket{1},\ket{0})\quad&\rightarrowtail\quad
T^{(1,1,1)}(\ket{1}\otimes\ket{0}\otimes\ket{0})=
                          \ket{1}\otimes\ket{0}\otimes\ket{0} \\
(\ket{1},\ket{1})\quad&\rightarrowtail\quad
T^{(1,1,1)}(\ket{1}\otimes\ket{1}\otimes\ket{0})=
                            \ket{1}\otimes\ket{1}\otimes\ket{1}
\end{align*}

One immediately realizes the difference with respect to the
classical case. The classical truth-table represents a typical
irreversible transformation:

\begin{align*}
(0,0)\quad&\rightarrowtail\quad 0 \\
(0,1)\quad&\rightarrowtail\quad 0 \\
(1,0)\quad&\rightarrowtail\quad 0 \\
(1,1)\quad&\rightarrowtail\quad 1
\end{align*}

The arguments of the function determine the value, but not the
other way around. As is well known, irreversibility generally
brings about dissipation of information. Mathematically, however,
any Boolean function $f:\{0,1\}^n\mapsto~\{0,1\}^m$ can be
transformed into a {\it reversible\/} function
$\hat{f}:\{0,1\}^{n}\times \{0,1\}^{m}\mapsto \{0,1\}^{n}\times
\{0,1\}^{m}$ in the following way: $$ \forall u \in \{0,1\}^n \,\,
\forall v \in \{0,1\}^m: \hat{f}((u,v))=(u,v\oplus f(u)), $$ where
$\oplus$ is the sum modulo 2 pointwise defined. The function that
is obtained by making reversible the irreversible classical
``and'' corresponds to the Toffoli gate. The classical ``and'' is
then recovered by fixing the third input bit to 0.

Accordingly, the three arguments $(0,0)$, $(0,1)$, $(1,0)$ turn
out to correspond to three distinct values, represented by the
triples $(0,0,0)$, $(0,1,0)$, $(1,0,0)$. The price we have paid in
order to obtain a reversible situation is the increasing of the
complexity of our Hilbert space. The function ${\tt AND}$
associates to pairs of arguments, belonging to the two-dimensional
space $\c2$, values belonging to the space $\otimes^3\c2$ (whose
dimension is $2^3$).

All this happens in the simplest situation, when one is only
dealing with elements of the basis (in other words, with precise
pieces of information). Let us examine the case where the function
{\tt AND} is applied to arguments that are superpositions of the
basis-elements in the space $\c2 $. Consider the following {\it
qubit\/} pair: $$ \ket{\psi}=a_0\ket{0}+a_1\ket{1}\,,\,\,
\ket{\varphi}=b_0\ket{0}+b_1\ket{1}. $$ By applying the
definitions of ${\tt AND}$ and of $T^{(1,1,1)}$, we obtain:

$$ {\tt AND}(\ket{\psi},\ket{\varphi})= a_1b_1\ket{1,1,1}+
a_1b_0\ket{1,0,0}+
 a_0b_1\ket{0,1,0}+ a_0b_0\ket{0,0,0}.
$$

This result suggests a quite natural logical interpretation. The
four basis-elements that occur in our superposition-vector
correspond to the four cases of the truth-table for the classical
conjunction: $$ (1,1,1), (1,0,0), (0,1,0), (0,0,0). $$
 However here, differently from the classical situation,  each case is accompanied by a complex number, which represents a characteristic quantum {\it amplitude\/}. By applying the ``Born rule" we will obtain the following interpretation:
$|a_1b_1|^2$ represents the probability-value that both the {\it
qubit\/}-arguments are equal to $\ket{1}$, and consequently their
conjunction is $\ket{1}$. Similarly in the other three cases.

The logical gate $\tt AND$ refers to a very special situation,
characterized by a Hilbert space having the form $\otimes^3\D
C^2$. However, our procedure can be easily generalized. The
Toffoli gate can be defined in any Hilbert space having the form:
$$ (\tensorn)\otimes(\tensorm)\otimes\c2 (=\otimes^{n+m+1}\D C^2).
$$

\begin{definition}
{\it The Toffoli gate\/} $T^{(n,m,1)}$
\label{de:toffoligenerale}\nl The {\it Toffoli gate \/}
$T^{(n,m,1)}$ is the linear operator $$ T^{(n,m,1)}:
(\tensorn)\otimes(\tensorm)\otimes\c2\,\, \mapsto \,\,
(\tensorn)\otimes(\tensorm)\otimes\c2, $$ that is defined for any
element
$\ket{x_1,\ldots,x_n}\otimes\ket{y_1,\ldots,y_m}\otimes\ket{z}$ of
the computational basis of $\otimes^{n+m+1}\D C^2$ as follows: $$
T^{(n,m,1)}(\ket{x_1,\ldots,x_n}\otimes\ket{y_1,\ldots,y_m}\otimes\ket{z})
=\ket{x_1,\ldots,x_n}\otimes\ket{y_1,\ldots,y_m}\otimes\ket{\text{min}(x_n,y_m)
  \oplus z},
$$ where $\oplus$ represents the sum modulo $2$.
\end{definition}

On this basis one can immediately generalize our definition of
${\tt AND}$.

\begin{definition}
{\tt AND} \nl For any $\ket{\varphi}\in\tensorn$ and any
$\ket{\psi}\in\tensorm$: $$ {\tt
AND}\parto{\ket{\varphi},\ket{\psi}}:
=T^{(n,m,1)}\parto{\ket{\varphi}\otimes\ket{\psi}\otimes\ket{0}}.
$$
\end{definition}

How to deal in this context with the concept of negation? A
characteristic of quantum computation is the possibility of
defining a plurality of negation-operations: some of them
represent good generalizations of the classical negation. We will
first consider a function $\tt NOT$ that simply inverts the value
of the last elements of any basis-vector. Thus, if
$\ket{x_1,\ldots,x_n}$ is any vector of the computational basis of
$\tensorn$, the result of the application of $\tt NOT$ to
$\ket{x_1,\ldots,x_n}$ will be $\ket{x_1,\ldots,1-x_n}$.

Consider first the simplest case, concerning the negation of a
single {\it qubit\/}. In such a case, the function ${\tt NOT}$
will be a unary function assuming arguments in the space $\c2$ and
values in the space $\c2$.

\begin{definition}
${\tt NOT}^{(1)}$ \label{de:not} \nl For any
$\kphi=a_0\kz+a_1\ko\in \c2$: $$ {\tt
NOT}^{(1)}\parto{\kphi}:=a_1\kz+a_0\ko. $$
\end{definition}

One can immediately check that ${\tt NOT}$ represents a good
generalization of the classical truth-table. Consider the
basis-elements $\kz$ e $\ko$. In such a case we will obtain:
\begin{align*}
{\tt NOT}^{(1)}(\ko)=\kz; \\ {\tt NOT}^{(1)}(\kz)=\ko.
\end{align*}
The quantum logical gate ${\tt NOT}^{(1)}$ can be easily
generalized in the following way.

\begin{definition}
${\tt NOT}^{(n)}$ \label{de:notgenerale} \nl

${\tt NOT}^{(n)}$ is the map $$ {\tt
NOT}^{(n)}:\,\tensorn\mapsto\,\tensorn $$ s.t. for any
$\kpsi=\sumn a_j\kj\in\tensorn$: $$ \NOT(\kpsi):=\sumn
a_j\ket{x_{j_1},\ldots,x_{j_{n-1}},1-x_{j_n}} $$

\end{definition}

The matrix corresponding to ${\tt NOT}^{(1)}$ will
be:

$$
\begin{pmatrix}
 0 & 1 \cr 1 & 0 \cr
\end{pmatrix}
$$ The matrix corresponding to ${\tt NOT}^{(n)}$ will be the
following $2^n\times 2^n$ matrix: $$
\begin{pmatrix}
 0 & 1  &0 &0 &. &. &. &0 \cr
 1 & 0  &0 &0 &. &. &. &0 \cr
 0 & 0  &0 &1 &0 &. &. &0 \cr
  0 & 0  &1 &0 &0 &. &. &0 \cr
   .& .  &. &. &. &. &. &. \cr
    .& .  &. &. &. &. &. &. \cr
0 &. &. &. &0 &0 &0 &1 \cr
   0 &.  &. &. &0 &0 &1 &0 \cr
 \cr
\end{pmatrix}
$$

We will omit the index $n$ in $\NOT^{(n)}$ if  no confusion is
possible.

Finally, how to introduce a reasonable disjunction? A gate $\tt
OR$ can be naturally defined in terms of $\tt AND$ and $\tt NOT$
via de Morgan.

\begin{definition} {\em $\tt OR$}

\label{de:demorgan} For any $\kphi\in\tensorn$ and
$\kpsi\in\tensorm$: $$
\OR(\kphi,\kpsi)=\NOT\parto{\AND\parto{\NOT({\kphi}),\NOT({\kpsi})}}.
$$

\end{definition}

The quantum logical gates we have considered so far are, in a
sense, ``semiclassical". A quantum logical behaviour only emerges
in the case where our gates are applied to superpositions. When
restricted to classical registers, our gates turn out to behave as
classical truth-functions. We will now investigate {\em genuine
quantum gates\/} that may transform classical registers into
quregisters that are superpositions.

One of the most significant genuine quantum gates is the {\em
squareroot of the negation $\NOT$}, which will be indicated by
$\SQR$. As suggested by the name, the characteristic property of
the gate $\SQR$ is the following: for any quregister $\kpsi$,
$$\SQR \SQR \kpsi= \NOT \kpsi.$$

In other words: applying twice the squareroot of the negation
``means" negating.

Interestingly enough, the gate $\SQR$ has some interesting
physical models (and implementations). As an example, consider an
idealized atom with a single electron and two energy levels: a
{\em ground state\/} (identified with $\ket{0}$) and an {\em
excited state\/} (identified with $\ket{1}$). By shining a pulse
of light of appropriate intensity, duration and wavelength, it is
possible to force the electron to change energy level. As a
consequence, the state (bit) $\ket{0}$ is transformed into the
state (bit) $\ket{1}$, and viceversa: $$\ket{0}\rightarrowtail
\ket{1}; \,\,\,\ket{1}\rightarrowtail \ket{0}. $$

We have obtained a typical physical model for the gate $\NOT$.
Now, by using a light pulse of half the duration as the one needed
to perform the $\NOT$ operation, we effect a half-flip between the
two logical states. The state of the atom after the half pulse is
neither $\ket{0}$ nor$\ket{1}$, but rather a superposition of both
states. As observed by Deutsch, Ekert, Lupacchini (\cite{DEL}):
\begin{quote}
Logicians are now entitled to propose a new logical operation
$\SQR$. Why? Because a faithful physical model for it exists in
nature.
\end{quote}

Interestingly enough, the gate $\SQR$ seems to have also some
linguistic ``models". For instance, consider the French language.
Put: $$\SQR =\,\,\,\text{``ne"}= \,\,\,\text{``pas"}.$$ We will
obtain: $$\SQR \SQR = \,\,\,\text{``ne....pas"}\,\,\,=\NOT.$$

Let us now give the mathematical definition of $\SQR$. We will
first consider the simplest case, which refers to the space $\c2$.

\begin{definition}
$\SQR^{(1)}$ \label{de:radicenegata}
\nl $\SQR^{(1)}$ is the map $$
\SQR^{(1)}:\,\c2\to\,\c2 $$ such that for any
$\kpsi=:a_0\kz+a_1\ko$:
 $$
\SQR^{(1)}(\kpsi):=\dfrac{1}{2}\parq{(1+i)a_0+(1-i) a_1}\kz +
\dfrac{1}{2}\parq{(1-i) a_0+(1+i) a_1}\ko,
$$

 where $i$ is the imaginary
unit.
\end{definition}

It turns out that the matrix associated to $\SQR^{(1)}$ is
$$
\begin{pmatrix}
\frac{1}{2}+\frac{i}{2} &\frac{1}{2}-\frac{i}{2} \\
{}  &{}\\
\frac{1}{2}-\frac{i}{2} & \frac{1}{2}+\frac{i}{2}
\end{pmatrix}
$$ Thus, $\SQR^{(1)}$ transforms the two bits $\kz$ and $\ko$ into
the superposition states $\dfrac{1}{2}(1+i) \kz +
\dfrac{1}{2}(1-i) \ko$ and $\dfrac{1}{2}(1-i) \kz +
\dfrac{1}{2}(1+i) \ko$, respectively.

The quantum logical gate $\SQR^{(1)}$ can be easily generalized in the following way.
\begin{definition}
$\SQR^{(n)}$
\nl $\SQR^{(n)}$ is the map $$
\SQR^{(n)}:\,\tensorn\mapsto\,\tensorn $$ such that for any
$\kpsi=\sumn a_j\kj\in\tensorn$:
$$ \SQR^{(n)}(\kpsi):=\sumn
a_j\ket{x_{j_1},\ldots,x_{j_{n-1}}}\otimes\SQR^{(1)}(\ket{x_{j_n}})
$$
\end{definition}

It is easy to see that for any $n$, $\SQR^{(n)}$ is a unitary
operator such that $$ \SQR^{(n)}\SQR^{(n)}=\NOT^{(n)}. $$ The
matrix associated to the quantum logical gate $\SQR$ is the
$(2^n)\times (2^n)$ matrix of the form
$$
\dfrac{1}{2}\begin{pmatrix}
1+i &1-i  &. &. &. &. &. &. &.\\
1-i &1+1  &. &. &. &. &. &. &.\\
. &.  &1+i &1-i &. &. &. &. &.\\
. &.  &1-i &1+i &. &. &. &. &.\\
. &. &. &. &. &. &. &. &.\\
. &. &. &. &. &. &. &. &.\\
. &. &. &. &. &. &. &1+i &1-i\\
. &. &. &. &. &. &. &1-1 &1+i\\
\end{pmatrix}
$$
We will omit the index $n$ in $\SQR^{(n)}$ if  no confusion is
possible.
\begin{theorem}
\label{th: commutatore}
For any $n,m$ the following properties hold:
\begin{enumerate}\item[]
   \begin{enumerate}
  \item[(i)]\quad
  $T^{(n,m,1)}\SQR^{(n+m+1)}=\SQR^{(n+m+1)}T^{(n,m,1)}$;
  \item[(ii)]\quad $\SQR^{(n)}\NOT^{(n)}=\NOT^{(n)}\SQR^{(n)}$.
\end{enumerate}
  \end{enumerate}
\end{theorem}


\section{The probabilistic content of the quantum logical gates}
For any quregister one can define a natural {\em
probability-value\/}, which will play an important role in our
quantum computational semantics.

Suppose a vector $$ \ket{\varphi}=\sumn a_j\ket{j} \in \tensorn.$$
Let us first define two particular sets of coefficients that occur
in the superposition-vector $\varphi$:$$
\cpphi:=\parg{a_j\,:\,1\le j\le 2^n-1 \ \text{and $j$ is odd}}, $$

$$ \cmphi:=\parg{a_j\,:\,1\le j\le 2^n-1 \ \text{and $j$ is
even}}\cup \parg{0}. $$

Clearly, the elements of $\cpphi$ ($\cmphi$) represent the {\it
amplitudes\/} associated to the different vector-basis of
$\tensorn$ ending with $1$ ($0$, respectively). Thus, $$
\cpphi=\parg{a_j\,:\, \ket{x_j}=
                 \para{x_{j_1}, \ldots, x_{j_{n-1}}, 1}},
$$ and $$ \cmphi=\parg{a_j\,:\, \ket{x_j}=
                 \para{x_{j_1}, \ldots, x_{j_{n-1}}, 0}}.
$$ On this basis, we can now define the probability-value of any
vector having length less than or equal to 1. \nl

\begin{definition}{\em The probability-value of a vector}
 \label{de:registroprobabile} \nl
 Let
$\ket{\psi}=\sumn a_j\ket{j}$ be any vector of $\tensorn$ such
that $\sumn |a_j|^2\le 1$. Then the probability-value of
$\ket{\psi}$ is defined as follows: $$ \P(\kpsi):=\sum_{a_j \in
C^+\ket\psi} |a_j|^2. $$
\end{definition}

According to our definition, in order to calculate the
probability-value of a quregister $\ket{\psi}$ one has to perform
the following operations:
\begin{itemize}
\item \space consider all the amplitudes $a_j$ that are associated
to a basis-element ending with $1$;
\item \space take the squared modules $|a_j|^2$ of all these
complex numbers $a_j$;
\item \space  sum  all the real numbers $|a_j|^2$.
\end{itemize}

One can prove:

\begin{lemma}  \label{le:ortogonale} \nl
\begin{enumerate}\item[]
   \begin{enumerate}
  \item[(i)]\space If  $\kpsi=\kphin$ is any unit vector of $\tensorn$, then
 $$ \sum_{a_j \in
C^+\ket\psi} |a_j|^2+\sum_{a_j \in C^-\ket\psi} |a_j|^2=1. $$
  \item[(ii)]\space Let $\kpsi=\sumn a_j\kj$ and $\kphi=\sumn
b_j\kj$ be any two orthogonal vectors of $\tensorn$ s.t.
$\norm{\kpsi+\kphi}\le 1$ and $\forall j (0\le j\le 2^n-1)$:
$a_jb_j=0$. Then
 $$
 \P(\kpsi+\kphi)=\P(\kpsi)+\P(\kphi).
 $$

\end{enumerate}
  \end{enumerate}

\end{lemma}

From an intuitive point of view, $\P(\kpsi)$ represents ``the
probability" that our quregister $\kpsi$ (which is a
superposition) ``collapses" into a classical register whose last
element is $1$.

The following theorem describes some interesting relations between
the probability function $\P$ and our basic logical gates.
\begin{theorem}\label{th:radice}
Let $\kpsi=\sumn a_j\kj$ and $\kphi=\summ b_k\kk$ be two unit
vectors of $\tensorn$ ($\tensorm$, respectively). The following
properties hold:
\begin{enumerate}\item[]
   \begin{enumerate}
  \item[(i)]\quad $\P(\AND(\kpsi,\kphi))=\P(\kpsi)\P(\kphi)$;
  \item[(ii)]\quad $\P(\NOT(\kpsi))=1-\P(\kphi)$;
  \item[(iii)]\quad $\P(\OR(\kpsi,\varphi))=\P(\kpsi)+\P(\kphi)-\P(\kpsi)\P(\kphi$;
    \item[(iv)]\quad $\P(\SQR(\kpsi))=\sumppsi\mid \dfrac{1}{2}(1-i)a_{j-1}
    +\dfrac{1}{2}(1+i)a_j\mid^2$.
 \item[(v)]\quad $\P(\SQR\,\NOT(\kpsi))=\P(\NOT\SQR(\kpsi))=\sumppsi\mid
 \dfrac{1}{2}(1+i)a_{j-1}+
    \dfrac{1}{2}(1-i)a_j\mid^2$.
      \item[(vi)]\quad $\P(\SQR(\AND(\kpsi,\kphi)))=\frac{1}{2}$.
\end{enumerate}
  \end{enumerate}
\end{theorem}

Condition (i) of Theorem \ref{th:radice} represents a quite
unusual property for probabilistic contexts: any pair of
quregisters seems to behave here like a classical pair of
independent events (such that the probability of their conjunction
is the product of the probabilities of both members). At the same
time, condition (ii) and (iii) appear to be well behaved with
respect to standard probability theory. As a consequence, we
obtain:
\begin{itemize}
\item \space differently from classical probability (and also from
standard quantum probability) $\AND$, $\OR$, $\NOT$ have a
``truth-functional behaviour" with respect to the function $\P$:
the probability of the ``whole" is determined by the probabilities
of the parts.
\item \space The gate $\SQR$ is not
truth-functional. It may happen at the same time that:
$\P(\kpsi)=\P(\kphi)$ and $\P(\SQR(\psi))\not=\P(\SQR(\kphi))$.
For example, let
$\kpsi:=\frac{\sqrt{2}}{2}\kz+\frac{\sqrt{2}}{2}\ko$ and
$\kphi:=\frac{\sqrt{2}}{2}\kz+\frac{\sqrt{2}}{2}
\parto{\frac{\sqrt{2}}{2}+\frac{\sqrt{2}}{2}i}\ko$.
Clearly, $\P(\kpsi)=\P(\kphi)=\frac{1}{2}$. However,
$\P(\SQR(\psi))=\frac{1}{2}$ and $\P(\SQR(\kphi))=\frac{1}{8} +
{\left( \frac{1}{2} - \frac{1}{2\,{\sqrt{2}}} \right) }^2\approx
0.146447$.
\end{itemize}

Both the operators $\NOT^{(1)}$ and $\SQR^{(1)}$ have a fixed point.
For instance, the vector
$\kpsi=\frac{1}{\sqrt{2}}\kz+\frac{1}{\sqrt{2}}\ko$ is a fixed
point of $\NOT^{(1)}$, since $\NOT^{(1)}(\kpsi)=\kpsi$. Clearly,
if $\psi$ is a fixed point of $\NOT^{(1)}$, then also
$\NOT^{(1)}(\kpsi$ is
 a fixed point of $\NOT^{(1)}$. At  the same time, the
 vector $\frac{e^{i\vartheta}}{\sqrt{2}}(\kz+\ko)$
 turns out to be  a fixed point of  $\SQR^{(1)}$.

\section{Quantum computational semantics}

The starting point of the quantum computational semantics is quite
different from the standard quantum logical approach. The meanings
of sentences are here represented by quregisters. From an
intuitive point of view, one can say that the meaning of a
sentence is identified with the {\it information quantity\/}
encoded by the sentence in question (where information is of
course
 measured by means of the quantum unit).

Consider a sentential language $\C L$ with the following
connectives: the {\it negation\/} ($\neg$), the {\it
conjunction\/} ($\curlywedge$) and the {\it square root of the
negation\/} ($\sqr$). The notion of {\it sentence\/} (or {\it
formula\/}) of $\C L$ is defined in the expected way. Let $Form^\C
L$ represent the set of all sentences of $\C L$. We will use the
following metavariables: $\bf p,q, r,\ldots$ for atomic sentences
and $\alpha $, $\beta $, $\gamma,\ldots$ for sentences. The
connective {\em disjunction\/} ($\curlyvee$) is supposed defined
via de Morgan's law: $$
  \alpha \curlyvee \beta := \neg \part{\neg \alpha \curlywedge \neg \beta}.
$$

We will now introduce the basic concept of our semantics, the
notion of {\it quantum computational realization\/}: an
interpretation of the language $\C L$, such that the meaning
associated to any sentence is a quregister. As a consequence, the
{\it space of the meanings \/} corresponds here to a
 variable Hilbert space (instead of a unique Hilbert space). Any space of
  this kind will be a product space $\otimes^n \c2$.

\begin{definition}  \label{de:realcomputo}
{\it Quantum computational realization\/} \nl A {\it quantum
computational realization\/} of $\C L$ is a function $Qub$
associating to any sentence $\alpha$ a quregister in a Hilbert
space $\tensorn$ (where $n$ is a function of the length of
$\alpha$): $$ Qub:\, Form^\C L\,\to \bigcup_n\tensorn. $$ We will
also write $\ket{\alpha}$ instead of $Qub(\alpha)$; and we will
call $\ket{\alpha}$ the {\it information-value\/} of $\alpha$. The
following conditions are required:
\begin{enumerate}
   \item[]
  \begin{enumerate}
\item[(i)]\space $\ket{\bf p}\in\c2$;
\item[(ii)]\space Let $\ket{\beta}\in \tensorn$.
Then \nl $\ket{\neg\beta}={\tt NOT}(\ket{\beta})\in \tensorn$;
\item[(iii)]\space Let  $\ket{\beta}\in \tensorn,\,\ket{\gamma}\in \tensorm$.
Then: \nl $\ket{\beta \curlywedge \gamma}={\tt
AND}(\ket{\beta},\ket{\gamma})\in
      (\tensorn)\otimes(\tensorm)\otimes \IC^2$;
     \item[(iv)]\space Let $\ket{\beta}\in \tensorn$.
Then \nl $\ket{\sqr\beta}={\SQR}(\ket{\beta})\in \tensorn$;
\end{enumerate}
  \end{enumerate}
\end{definition}
Our definition univocally determines, for any sentence $\alpha$,
the Hilbert space $\tensorn$ to which $\ket{\alpha}$ belongs.
Clearly, $n$ is a function of the number of all occurrences of
atomic sentences in $\alpha$. Since the meaning associated to a
given sentence reflects the logical form of the sentence in
question, we can say that our semantics has a typical {\it
intensional\/} character.

As we have seen, a characteristic of our semantics is to identify
the meanings
 of the linguistic sentences with unit vectors of variable Hilbert spaces.
As a consequence, we will obtain that the information-value of a
sentence naturally determines a probability-value for that
sentence.

Let $Qub$ be a quantum computational realization and let $\alpha$
be any sentence with associated meaning $\ket{\alpha}$. Like all
qubit-registers, also our $\ket{\alpha}$ will have a
probability-value, which (according to Definition
\ref{de:registroprobabile}), is determined as follows:

$$ \P(\ket{\alpha)}:=\sum_{a_j \in C^+\ket\alpha} |a_j|^2. $$

On this basis, one can naturally define the probability-value of
any sentence of our language:

\begin{definition}
{\it The probability-value of $\alpha$\/}
\label{de:valoreprobabile} $$ \P(\alpha):=\sum_{a_j \in
C^+\ket\alpha} |a_j|^2. $$
\end{definition}

As an example, let us first consider the simplest case, where
$\alpha$ is an atomic sentence; in this case, its
information-value will belong to the two-dimensional space $\c2$.
Suppose, for instance, that $\ket{\alpha}$ has the form: $$
a_0\ket{0}+a_1\ket{1}. $$

Then, the probability-value of $\alpha$ will be: $$
\P(\alpha)=|a_1|^2. $$

Thus, $\P(\alpha)=|a_1|^2$ represents the probability that our
{\it uncertain\/} information $\ket{\alpha}$ corresponds to the
{\it precise\/} information $\ket{1}$.

>From an intuitive point of view, our definition, clearly,
attributes a privileged role to one of the two basic {\it
qubits\/} (belonging to the basis of $\c2$): the {\it qubit\/}
$\ket{1}$. In such a way, $\ket{1}$ is dealt with as the
truth-value {\it True\/}.

Consider now the case of a molecular sentence $\alpha$. Its
information-value $\ket{\alpha}$ will belong to the space
$\tensorn$, where $n$ ($\ge 3$) depends on the length of $\alpha$.
The dimension of $\tensorn$ is $2^n$. Hence $\ket{\alpha}$ will
generally be a superposition of elements of the basis of
$\tensorn$. Thus, we will have:
$$ \ket{\alpha}=\sumn a_j\kj.
$$

>From the logical point of view, any $\kj$ (element of the basis of
$\tensorn$) represents a possible case of a ``reversibile
truth-table" for $\alpha$. For instance, suppose $\alpha$ has the
form $\mathbf p \curlyvee \mathbf q$, where: $$ \ket{\mathbf
p}=a_0\ket{0}+a_1\ket{1}, \,\, \ket{\mathbf
q}=b_0\ket{0}+b_1\ket{1}. $$ By applying the definitions of
quantum computational realization and of ${\OR}$, we will obtain:
$$ \ket{\mathbf p \curlyvee \mathbf q}= a_1b_1\ket{1,1,1}+
a_1b_0\ket{1,0,1}+ a_0b_1\ket{0,1,1}+ a_0b_0\ket{0,0,0}. $$

We know that the number $|a_1b_1|^2$ represents the probability
that both the members of our disjunction are true and that,
consequently, the disjunction is true. Similarly in the other
cases. In order to calculate the probability of the truth of
$\mathbf p \curlyvee \mathbf q$, it will be sufficient to sum the
three probability-values corresponding to the three cases where
the final result is {\it True\/} (that is the cases of the vectors
$\ket{1,1,1}$, $\ket{1,0,1}$, $\ket{0,1,1}$). On this basis, we
will be able to assign to the disjunction $\mathbf p \curlyvee
\mathbf q$ the following probability-value: $$ |a_1b_1|^2+
|a_1b_0|^2 +|a_0b_1|^2. $$

We can now define the notions of {\it truth\/}, {\it logical
truth\/}, {\it consequence \/} and {\it logical consequence \/}.

\begin{definition}
{\it Truth and logical truth} \label{de:verocompito} \nl A
sentence $\alpha$ is {\it true in a realization\/} $Qub$
($\models_{Qub}\alpha$) iff $\P(\alpha)=1$.\nl $\alpha$ is a {\em
logical truth\/} ($\models \alpha$) iff for any realization $Qub$,
$\models_{Qub} \alpha$.
\end{definition}

\begin{definition}
\label{de:conseguendocompiti} {\it Consequence and logical
consequence\/} \nl $\beta$ is a {\it consequence of $\alpha$ in
the realization\/} $Qub$ ($\alpha\models_{Qub}\beta$) iff
$\P(\alpha)\le \P(\beta)$; \nl $\beta$ is a {\it logical
conseguence\/} of $\alpha$ ($\alpha\models\beta$) iff for any
$Qub$: $\alpha\models_{Qub}\beta$.
\end{definition}

Let us call the logic characterized by this semantics {\it quantum
computational logic\/} ({\bf QCL}).

Some interesting examples of logical consequences that hold in
{\bf QCL} are the following:
\begin{theorem}
\nl
\begin{enumerate}\item[]
   \begin{enumerate}
  \item[(i)]\space $\alpha\models\neg\neg\alpha,\quad \neg\neg\alpha\models\alpha$;
  \\ (double negation)
  \item[(ii)]\space $\sqr\sqr\alpha\models\neg\alpha,\quad
  \neg\alpha\models\sqr\sqr\alpha$;
  \item[(iii)]\space $\alpha\curlywedge\beta\models\beta\curlywedge\alpha,
                \quad \alpha\curlyvee\beta\models\beta\curlyvee\alpha$;
                \\(commutativity)
     \item[(iv)]\space $\alpha\curlywedge (\beta\curlywedge\gamma)\models
                        (\alpha\curlywedge\beta)\curlywedge\gamma,
        \quad  (\alpha\curlywedge\beta)\curlywedge\gamma\models \alpha\curlywedge
        (\beta\curlywedge\gamma)$;
        \\ (associativity)
    \item[(v)] \space   $\alpha\curlyvee(\beta\curlyvee\gamma)\models
                        (\alpha\curlyvee\beta)\curlyvee\gamma,
        \quad  (\alpha\curlyvee\beta)\curlyvee\gamma\models \alpha\curlyvee
        (\beta\curlyvee\gamma)$;
  \\           (associativity)
 \item[(vi)]\space $\neg(\alpha\curlywedge\beta)\models\neg\alpha\curlyvee
 \neg\beta,
                \quad  \neg\alpha\curlyvee\neg\beta\models  \neg(\alpha\curlywedge
                \beta)$;
                \\     (de Morgan)

  \item[(vii)]\space  $\neg(\alpha\curlyvee\beta)\models\neg\alpha\curlywedge
  \neg\beta,
                \quad  \neg\alpha\curlywedge\neg\beta\models  \neg(\alpha\curlyvee
                \beta) $
\\ (de Morgan)
\item[(viii)]\space $\alpha\curlywedge\alpha\models\alpha$.
\\ (semiidempotence 1)
\item[(ix)]\space $\alpha\curlywedge(\beta\curlyvee\gamma)\models(\alpha\curlywedge
\beta)\curlyvee(\alpha\curlywedge\gamma)$.
\\ (distributivity 1)
\end{enumerate}
  \end{enumerate}
\end{theorem}

Some logical consequences and some logical truths that are
violated in {\bf QCL} are the following:
\begin{theorem}
\nl
\begin{enumerate}\item[]
  \begin{enumerate}
\item[(i)]\space  $\alpha\not\models\alpha\curlywedge\alpha$;
\\ (semiidempotence 2)
\item[(ii)]\space $\not\models\alpha\curlyvee\neg\alpha$
\\ (excluded middle)
\item[(iii)]\space $\not\models\neg(\alpha\curlywedge\neg\alpha)$;
\\ (non contradiction)
\item[(iv)]\space $(\alpha\curlywedge\beta)\lor(\alpha\curlywedge\gamma)
\not\models
  \alpha\curlywedge(\beta\curlyvee\gamma)$.
\\ (distributivity)
\end{enumerate}
  \end{enumerate}
  \end{theorem}
\begin{proof}
(i)-(iii)\space Take
$\ket{\alpha}:=\frac{\sqrt{2}}{2}\kz+\frac{\sqrt{2}}{2}\ko$. Then,
$\P(\alpha)=\frac{1}{2}$, $\P(\alpha\mand\alpha)=\frac{1}{4}$,
$\P(\alpha\mor\neg\alpha)=\P(\neg(\alpha\mand\neg\alpha))=\frac{3}{4}$.
\nl (iv)\space Take
$\ket{\alpha}=\ket{\beta}:=\frac{\sqrt{2}}{2}\kz+\frac{\sqrt{2}}{2}\ko$
and $\ket{\gamma}:=\frac{\sqrt{3}}{2}\kz+\frac{1}{2}\ko$. Then
$\P((\alpha\mand\beta)\lor(\alpha\mand\gamma))=\frac{11}{12}>\frac{10}{12}=
 \P(\alpha\mand(\beta\mor\gamma))$.

\end{proof}

{\bf QCL\/} turns out to be a non standard form of quantum logic.
Conjunction and disjunction do not correspond to lattice
operations, because they are not generally idempotent. Differently
from the usual (sharp and unsharp) quantum logics, the weak
distributivity principle
($(\alpha\curlywedge\beta)\curlyvee(\alpha\curlywedge\gamma)\models
  \alpha\curlywedge(\beta\curlyvee\gamma)$) breaks down.
  At the same time, the strong distributivity
  ($\alpha\curlywedge(\beta\curlyvee\gamma)\models
  (\alpha\curlywedge\beta)\curlyvee(\alpha\curlywedge\gamma)
$), that is violated in orthodox quantum logic, is here valid.
Both the excluded middle and the non contradiction principles are
violated: as a consequence, we have obtained an example of an {\it
unsharp logic\/}.

The axiomatizability of {\bf QCL} is an open problem.

\providecommand{\bysame}{\leavevmode\hbox
to3em{\hrulefill}\thinspace}

\end{document}